\documentclass[preprint,12pt]{elsarticle}

\usepackage{amssymb}
\usepackage{axodraw}

\journal{Nuclear Physics B}

\begin{document}
\newcommand{\bqa}{\begin{eqnarray}}
\newcommand{\eqa}{\end{eqnarray}}
\newcommand{\nl}{\nonumber \\}
\newcommand{\als}{\alpha_S}
\newcommand{\dbar}{\bar D}
\newcommand{\qbar}{\bar q}

\begin{frontmatter}



\title{QCD corrections to $H \to gg$ in FDR}


\author{Roberto Pittau}

\address{Departamento de F\'isica Te\'orica y del Cosmos and CAFPE,
Campus Fuentenueva s. n., Universidad de Granada, E-18071 Granada, Spain\\
e-mail: pittau@ugr.es
}

\begin{abstract}
I apply FDR -a recently introduced Four Dimensional Regularization approach to quantum field theories- to compute the NLO QCD corrections to $H \to gg$ in the large top mass limit. The calculation involves all key ingredients of massless QCD, namely ultraviolet, infrared and collinear divergences, besides $\als$ renormalization.
I show in detail how the correct result emerges in FDR, and discuss the translation rules to dimensional regularization.
\end{abstract}


\end{frontmatter}


\section{Introduction}
\label{intro}
Many of the difficulties of higher order calculations in quantum field theory (QFT) can be traced back to the treatment, in the framework of dimensional regularization (DR)~\cite{'tHooft:1972fi}, of the infinities arising in the intermediate steps of the computation.
Ultraviolet (UV), infrared (IR) and collinear (CL) divergences are first dimensionally regulated, and then renormalized away -in the UV case- or canceled  by combining virtual and real contributions, or reabsorbed in the collinear behavior of the initial state parton densities.
In order to attack the problem numerically, it is often necessary to subtract and add back approximations of the IR/CL singular structures.  At one loop, several well tested subtraction procedures have been introduced in the last two decades~\cite{Frixione:1995ms,Catani:1996vz,Kosower:1997zr,Campbell:1998nn,Catani:2002hc,Nagy:2003qn}. At two loops and beyond, the situation is more involved, but progress is under way~\cite{Binoth:2000ps,Anastasiou:2003gr,Binoth:2004jv,GehrmannDeRidder:2005cm,Catani:2007vq,Czakon:2010td}.

The first obvious ingredient, which may lead to a significant simplification in the above picture, is a computational procedure in which all parts of the calculation can be directly treated in four dimensions. As for the virtual contribution, the FDR approach has been recently introduced in reference~\cite{Pittau:2012zd}, which allows a subtraction of the UV divergences at the level of the integrand, leaving a four dimensional integration over the loop momenta. In the same work, the use of FDR as an IR/CL regulator in QED is also suggested. 

In this paper, I present the first application of the FDR ideas in the context of massless QCD, where the issues related to gauge invariance are much more subtle than in the QED case. I concentrate, in particular, on the calculation of the ${\cal O}(\als)$ gluonic corrections to the $H \to gg$ decay in the $m_{top} \to \infty$ limit, and re-derive the well known fully inclusive result~\cite{Djouadi:1991tka,Larin:1995sq}
\bqa
\label{eq:eq1}
\Gamma(H \to gg) = \Gamma^{(0)}(\als(M_H^2)) 
\left[ 
1+\frac{95}{4}\,\frac{\als}{\pi}
\right]\,,
\eqa
where 
\bqa
\label{eq:gamma0}
\Gamma^{(0)}(\als(M_H^2))= \frac{G_F \als^2(M_H^2)}{36 \sqrt{2} \pi^3} M^3_H
\eqa
is the lowest order contribution, with $N_F= 0$ in $\als(M_H^2)$, since only gluons are considered.

 Despite its simplicity, all key ingredients of massless QCD are present in this process, such as the simultaneous occurrence of IR/CL divergences and UV renormalization. The fact that the correct expression is reproduced, shows that FDR is a valid and consistent calculational scheme in massless QFTs, and gives confidence on its potential to simplify multi-leg/loop computations.

The outline of the paper is as follows. Section~\ref{model} provides the set-up of  the calculation.  In Section~\ref{fdr}, I review the FDR treatment of the UV divergences and discuss its interplay with the IR and CL infinities. Section~\ref{virtual} presents the FDR computation of the virtual part, while Section~\ref{realtot} deals with the real contribution and its merging with the one-loop piece. The connection between FDR and DR is discussed in Section~\ref{fdrvsdr} and the final conclusions are drawn in Section~\ref{conc}.

\section{The model for $H \to gg$}
\label{model}
\begin{figure}[ht]
\begin{center}
\begin{picture}(440,200)(0,0)

\SetScale{0.75}

\SetOffset(27,150)
\Gluon(-30,35)(10,0){3}{6}
\Gluon(10,0)(50,35){3}{6}
\GlueArc(10,11)(15,20,160){3}{5}
\DashLine(10,0)(10,-35){3}
\Text(10,-37)[t]{$V_1$}

\SetOffset(107,150)
\Gluon(-30,35)(10,23){3}{5}
\Gluon(10,23)(50,35){3}{5}
\GlueArc(10,11)(11,90,450){3}{9}
\DashLine(10,-4)(10,-35){3}
\Text(10,-37)[t]{$V_2$}

\SetOffset(187,150)
\Gluon(-30,35)(10,0){3}{6}
\GlueArc(-3,11)(15,-28,125.8){3}{5}
\Gluon(10,0)(50,35){3}{6}
\DashLine(10,0)(10,-35){3}
\Text(10,-37)[t]{$V_3$}

\SetOffset(267,150)
\Gluon(-30,35)(10,0){3}{6}
\Gluon(10,0)(50,35){3}{6}
\GlueArc(23,11)(15,54,207){3}{5}
\DashLine(10,0)(10,-35){3}
\Text(10,-37)[t]{$V_4$}

\SetOffset(347,150)
\Gluon(-30,-12)(10,0){-3}{5}
\Gluon(10,0)(50,-12){-3}{5}
\GlueArc(10,11)(11,-90,270){3}{9}
\DashLine(10,2)(10,-35){3}
\Text(10,-37)[t]{$V_5$}

\SetOffset(67,50)
\Gluon(-30,35)(10,0){3}{6}
\GCirc(-10,19){8}{0.6}
\Gluon(10,0)(50,35){3}{6}
\DashLine(10,0)(10,-35){3}
\Text(10,-37)[t]{$V_6$}

\SetOffset(147,50)
\Gluon(-30,35)(10,0){3}{6}
\Gluon(10,0)(50,35){3}{6}
\GCirc(30,19){8}{0.6}
\DashLine(10,0)(10,-35){3}
\Text(10,-37)[t]{$V_7$}

\SetOffset(227,50)
\LongArrow(-30,25)(-10,8)
\Text(-12,8)[tr]{$p_i$}
\Gluon(-30,35)(10,0){3}{6}
\Gluon(-13,24)(8,44){3}{3}
\LongArrow(50,25)(30,8)
\Text(40,8)[tr]{$p_k$}
\LongArrow(15,40)(2,28)
\Text(15,35)[tl]{$p_j$}
\Gluon(10,0)(50,35){3}{6}
\DashLine(10,0)(10,-35){3}
\Text(10,-37)[t]{$R_1(p_i,p_j,p_k)$}

\SetOffset(307,50)
\LongArrow(-30,25)(-10,8)
\Text(-12,8)[tr]{$p_1$}
\Gluon(-30,35)(10,0){3}{6}
\LongArrow(50,25)(30,8)
\Text(40,8)[tr]{$p_3$}
\Gluon(10,0)(50,35){3}{6}
\LongArrow(17,44)(17,22)
\Text(15,38)[tl]{$p_2$}
\Gluon(10,44)(10,4){-3}{4}
\DashLine(10,0)(10,-35){3}
\Text(10,-37)[t]{$R_2$}

\end{picture}
\caption{\label{fig:fig1} Virtual and real diagrams contributing 
to $H \to g g (g)$ at ${\cal O}(\als^3)$. The gray blobs in $V_6$ and $V_7$ represent gluon wave function corrections and the dashed line stands for the Higgs field. $R_1(p_i,p_j,p_k)$ corresponds to three diagrams with permuted gluons.}
\end{center}
\end{figure}
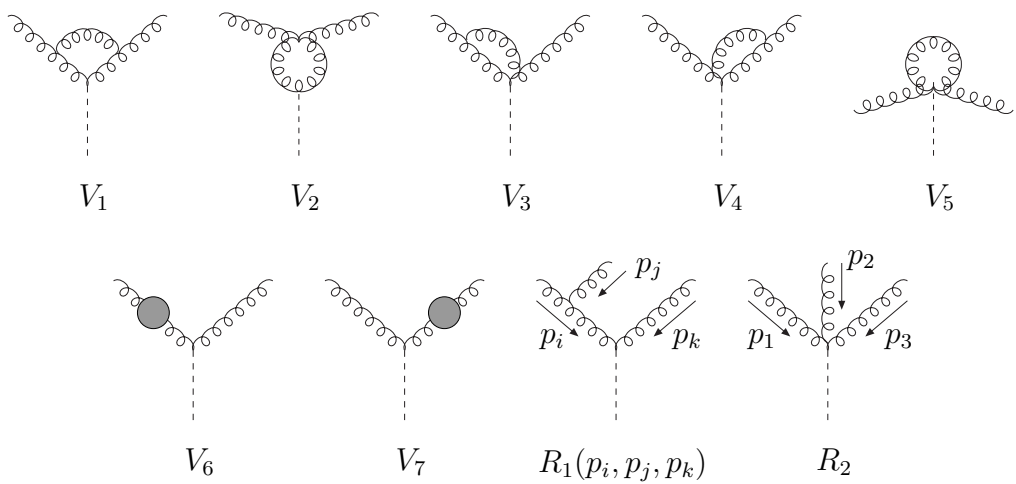

The effective interaction of one Higgs field $H$ with two, three and four gluons -mediated by an infinitely heavy top loop- is described by the Lagrangian~\cite{Shifman:1979eb,Dawson:1991au,Maltoni}
\bqa
{\cal L}_{\rm eff} &=& -\frac{1}{4} A H G^{a}_{\mu \nu} G^{a,\mu \nu}\,, 
\eqa
where
\bqa
\label{eq:A}
A &=& \frac{\als}{3\pi v}\left(1+\frac{11}{4}\frac{\als}{\pi} \right)
\eqa
and $v$ is the vacuum expectation value, $v^2= (G_F \sqrt{2})^{-1}$.
The corresponding Feynman rules are given in~\cite{Kauffman:1996ix}, and the diagrams for the decay rate $\Gamma(H \to gg)$ are drawn in Figure~\ref{fig:fig1}.
There are five graphs contributing to the virtual part  $\Gamma_{V}$ -without counting gluon wave function corrections- and four diagrams for the real radiation $\Gamma_{R}$. In the following, I separately compute, in FDR, the two pieces,
showing how IR/CL divergences drop in the sum
\bqa
\label{eq:eq2}
\Gamma_{V}(H \to gg) + \Gamma_{R}(H \to ggg)\,.  
\eqa

\section{FDR vs infinities}
\label{fdr}
The FDR subtraction of UV infinities is better illustrated with an explicit example. Consider the one-loop quadratically divergent rank-two tensor
\bqa
\label{eq:quadtens}
 \int d^4q \frac{q_\alpha q_\beta}{D_0 D_1}\,,
\eqa
with
\bqa
\label{defd}
D_i ~=~ q^2-d_i\,,~~~d_i ~=~ M^2_i-p^2_i- 2 (q\cdot p_i)\,,~~~p_0=0\,.
\eqa
Its UV convergence can be improved by first deforming the propagators by a vanishing amount $\mu^2$
\bqa
\label{eq:deform}
D_i \to \dbar_i~=~D_i-\mu^2\,,
\eqa
and then by repeatedly using the identity
\bqa
\label{eq:id}
	\frac{1}{\dbar_i}  = \frac{1}{\qbar^2}
		\Bigg(1+\frac{d_i}{\dbar_i} \Bigg)	\,,
\eqa
where
\bqa
\label{qdeformation}
\qbar^2~=~q^2-\mu^2\,.
\eqa
Note that the propagator deformation in eq.~(\ref{eq:deform}) is needed to avoid possible infrared divergences in the r.h.s. of eq.~(\ref{eq:id}). 
The integrand in eq.~(\ref{eq:quadtens}) can then be rewritten as
\bqa
\label{eq:exquadtens}
\frac{q_\alpha q_\beta}{\dbar_0 \dbar_1}= q_\alpha q_\beta \left(
 \left[\frac{1}{\qbar^4} \right]
+\left[\frac{d_0+d_1}{\qbar^6} \right]
+\left[\frac{d_1^2}{\qbar^8} \right]
+\frac{d_1^3}{\qbar^8 \dbar_1}
+ \frac{d_0 d_1}{\qbar^6 \dbar_1}
+ \frac{d_0^2}{\qbar^4 \dbar_0\dbar_1}
\right),
\eqa
where the terms in square brackets are UV divergent, but depend only on $\mu^2$.
The FDR {\em definition} of the integral in eq.~(\ref{eq:quadtens}) is obtained by integrating the expansion in eq.~(\ref{eq:exquadtens}), after dropping the divergent pieces, and taking the limit $\mu \to 0$:
\bqa
\label{eq:fdrquadtens}
B_{\alpha\beta}(p_1^2,M_0^2,M_1^2) &=&  \int [d^4q] \frac{q_\alpha q_\beta}{\dbar_0 \dbar_1} \nl
&\equiv&
\lim_{\mu \to 0} \int d^4q \,
q_\alpha q_\beta \left(
 \frac{d_1^3}{\qbar^8 \dbar_1}
+ \frac{d_0 d_1}{\qbar^6 \dbar_1}
+ \frac{d_0^2}{\qbar^4 \dbar_0\dbar_1}
\right)\,.
\eqa
The r.h.s. of eq.~(\ref{eq:fdrquadtens}) corresponds to a well defined
four dimensional integral, in which all UV divergences are explicitly
subtracted. Furthermore, IR and CL divergences get also regulated by the propagator deformation.
The gauge invariance properties of this definition are discussed in detail in~\cite{Pittau:2012zd} and~\cite{Donati:2013iya}. In the rest of this Section, I mostly concentrate on CL and IR infinities, and, in particular, on the matching between virtual and real contributions. 

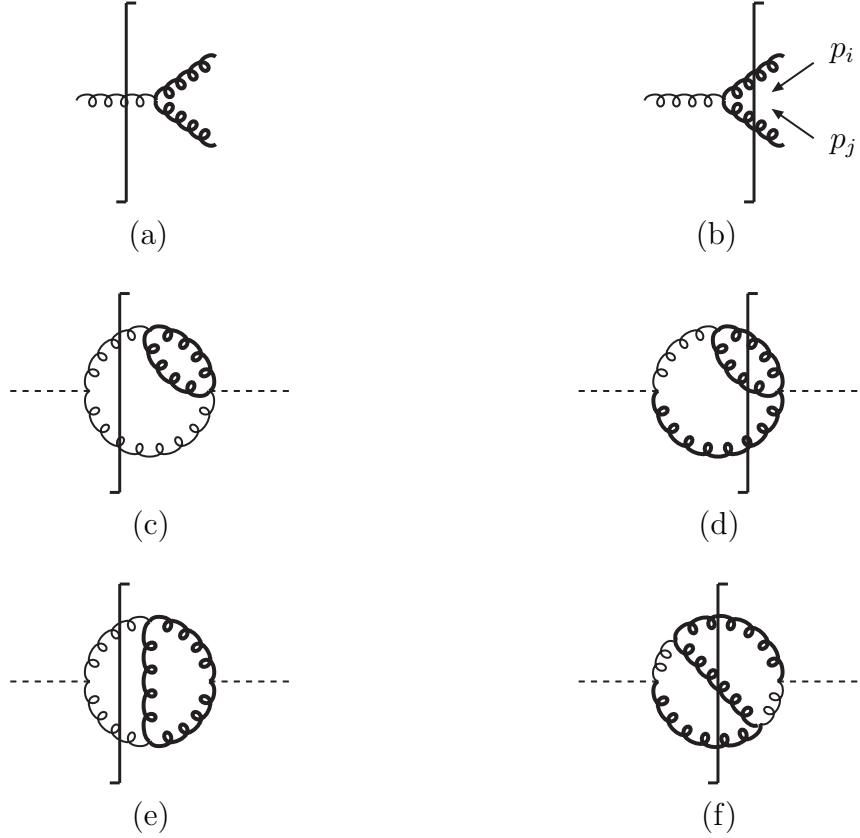
\begin{figure}[t]
\begin{center}
\begin{picture}(450,320)(0,0)

\SetScale{0.75}
\SetWidth{1}

\SetOffset(105,270)
\Gluon(-60,11)(-20,11){3}{4}
\SetWidth{2}
\Gluon(-20,11)(10,33){3}{4}
\Gluon(-20,11)(10,-11){-3}{4}
\SetWidth{1.5}
\Line(-35,60)(-30,60)
\Line(-35,-40)(-35,60)
\Line(-40,-40)(-35,-40)
\Text(-18,-37)[t]{(a)}
\SetWidth{1}

\SetOffset(320,270)
\Gluon(-60,11)(-20,11){3}{4}
\SetWidth{2}
\Gluon(-20,11)(10,33){3}{4}
\Gluon(-20,11)(10,-11){-3}{4}
\SetWidth{1}
\LongArrow(25,30)(5,16)
\LongArrow(25,-8)(5,6)
\Text(25,30)[lt]{$p_i$}
\Text(25,-13)[lb]{$p_j$}
\SetWidth{1.5}
\Line(-5,60)(0,60)
\Line(-5,-40)(-5,60)
\Line(-10,-40)(-5,-40)
\Text(-18,-37)[t]{(b)}

\SetOffset(80,160)
\SetWidth{1.5}
\Line(-5,60)(0,60)
\Line(-5,-40)(-5,60)
\Line(-10,-40)(-5,-40)
\SetWidth{1}
\DashLine(-60,11)(-20,11){3}
\SetWidth{2}
\GlueArc(40,41)(30,179,270){3}{4}
\GlueArc(10,11)(30,0,90){3}{4}
\SetWidth{1}
\GlueArc(10,11)(30,90,180){3}{4}
\GlueArc(10,11)(30,180,360){3}{8}
\DashLine(40,11)(80,11){3}
\Text(8,-37)[t]{(c)}

\SetOffset(295,160)
\SetWidth{1.5}
\Line(25,60)(30,60)
\Line(25,-40)(25,60)
\Line(20,-40)(25,-40)
\SetWidth{1}
\DashLine(-60,11)(-20,11){3}
\SetWidth{2}
\GlueArc(40,41)(30,179,270){3}{4}
\GlueArc(10,11)(30,0,90){3}{4}
\SetWidth{1}
\GlueArc(10,11)(30,90,180){3}{4}
\SetWidth{2}
\GlueArc(10,11)(30,180,360){3}{8}
\SetWidth{1}
\DashLine(40,11)(80,11){3}
\Text(8,-37)[t]{(d)}


\SetOffset(80,50)
\SetWidth{1.5}
\Line(-5,60)(0,60)
\Line(-5,-40)(-5,60)
\Line(-10,-40)(-5,-40)
\SetWidth{1}
\DashLine(-60,11)(-20,11){3}
\SetWidth{1}
\GlueArc(10,11)(30,90,180){3}{4}
\GlueArc(10,11)(30,180,270){3}{4}
\SetWidth{2}
\GlueArc(10,11)(30,0,90){3}{4}
\Gluon(10,-19)(10,41){3}{4}
\GlueArc(10,11)(30,270,360){3}{4}
\SetWidth{1}
\DashLine(40,11)(80,11){3}
\Text(8,-37)[t]{(e)}

\SetOffset(295,50)
\SetWidth{1.5}
\Line(10,60)(15,60)
\Line(10,-40)(10,60)
\Line(5,-40)(10,-40)
\SetWidth{1}
\DashLine(-60,11)(-20,11){3}
\SetWidth{1}
\GlueArc(10,11)(30,138,180){3}{2}
\GlueArc(10,11)(30,315,360){3}{2}
\SetWidth{2}
\GlueArc(10,11)(30,0,135){3}{6}
\Gluon(30,-11)(-11,33){3}{6}
\GlueArc(10,11)(30,180,315){3}{6}
\SetWidth{1}
\DashLine(40,11)(80,11){3}
\Text(8,-37)[t]{(f)}

\end{picture}
\caption{\label{fig:fig2} Gluon splitting IR/CL singularities regulated by massive (thick) gluons. The one-gluon cut in (a) contributes to the virtual part, the two-gluon cut in (b) to the real radiation. (c), (d), (e) and (f)  represent typical cut-diagrams contributing to $H \to gg(g)$.}
\end{center}
\end{figure}

A convenient starting point to study the CL singularities is the fully massless limit of eq.~(\ref{eq:fdrquadtens})
\bqa
\label{eq:b1}
B_{\alpha\beta}(0,0,0) = \lim_{\mu \to 0} \int d^4q \,
  \frac{q_\alpha q_\beta d_1^3}{\qbar^8 \dbar_1} =
 -8  p_1^\rho p_1^\sigma p_1^\tau \lim_{\mu \to 0} \int d^4q 
\frac{q_\alpha q_\beta q_\rho q_\sigma q_\tau}{\qbar^8 \dbar_1} = 0 \,,
\eqa
which vanishes, after tensor decomposition, since $p_1^2= 0$.
Analogously, one proves that
\bqa
\label{eq:b2}
B_{\alpha}(0,0,0) &=&  \int [d^4q] \frac{q_\alpha}{\dbar_0 \dbar_1} = 0\,, \nl
B(0,0,0) &=&  \int [d^4q] \frac{1}{\dbar_0 \dbar_1} = 0\,.
\eqa
Those results coincide with DR -in which scale-less integrals are zero- and are due to a cancellation between two $\ln(\mu^2)$ of CL and UV origin, respectively. For example
\bqa
\label{eq:bp2}
B(p^2,0,0) =  -i \pi^2 \lim_{\mu \to 0}\int_0^1 dx\, \left[
\ln(\mu^2 -p^2 x (1-x))-\ln(\mu^2) \right]\,, 
\eqa
where the first logarithm develops a CL singularity in the limit $p^2 \to 0$.
Thus, the virtual CL infinities, generated by $1 \to 2$ splittings of massless particles, are naturally regulated by the $\mu^2$-deformed propagators inside the loop, while the external momenta remain massless, as illustrated in Figures~\ref{fig:fig2} (a) and (c). The real counterpart of this procedure is exemplified in Figures~\ref{fig:fig2} (b) and (d), and corresponds to a phase space in which all the would be massless external particles are given a common mass $\mu$ and the internal ones stay massless. In other words, one has to replace~\footnote{See Figure~\ref{fig:fig2} (b).} 
\bqa
\label{eq:repl}
\frac{1}{2(p_i \cdot p_j)} \to \frac{1}{(p_i+p_j)^2}
\eqa
in any possible singular denominator of the real matrix element squared, integrate over the aforementioned massive phase space and take the limit $\mu \to 0$.

As for the IR divergences, the reasoning follows the same lines. For example, the only IR/CL divergent scalar one-loop three-point function generated by the cut in Figure~\ref{fig:fig2} (e) is~\footnote{The FDR integration corresponds to a normal integration, in this case, because $C(s)$ is UV finite.}
\bqa
\label{eq:1loopsc}
C(s)        &=& \int [d^4q] \frac{1}{\dbar_0 \dbar_1 \dbar_2}
= \lim_{\mu \to 0} \int d^4q \frac{1}{\dbar_0 \dbar_1 \dbar_2}\,, 
\eqa
with
\bqa
\label{eq:condi}
M_0^2 = M_1^2 = M_2^2 = p_1^2= p_2^2= 0\,,~~~s= -2(p_1 \cdot p_2)\,.
\eqa
By denoting
\bqa
\label{eq:mu0}
\mu_0=\frac{\mu^2}{s}\,,
\eqa
one computes
\bqa
\label{eq:cs}
C(s) = \lim_{\mu \to 0} \frac{i \pi^2}{2 s} \ln^2\Bigg(\frac{\sqrt{1-4\mu_0}+1}{\sqrt{1-4\mu_0}-1}\Bigg) =   
\frac{i \pi^2}{s} \left[
\frac{\ln^2(\mu_0)-\pi^2}{2} + i\,\pi \ln(\mu_0) \right],
\eqa
which is indeed fully matched by the inclusive real contribution in Figure~\ref{fig:fig2} (f), as will be shown in Section~\ref{fdrvsdr}.

In the following, I use the described approach to UV/CL/IR infinities to 
compute $\Gamma_{V}(H \to gg)$ and  $\Gamma_{R}(H \to ggg)$.

\section{The virtual part $\Gamma_{V}(H \to gg)$}
\label{virtual}
The calculation is greatly simplified by eqs.~(\ref{eq:b1}) and ~(\ref{eq:b2}). In fact, only diagrams $V_1$ and $V_2$ in Figure~\ref{fig:fig1} contribute -as in DR- and gluon wave function corrections vanish.
One computes
\bqa
\label{eq:gammav1}
\Gamma_{V}(H \to gg)=-3 \frac{\als}{\pi}\,\Gamma^{(0)}(\als)\,M^2_H\,
{\cal R}e \left[\frac{C(M^2_H)}{i \pi^2}\right]\,.
\eqa
This simple expression is obtained after a standard Passarino-Veltman~\cite{Passarino:1978jh} decomposition, the only subtlety being the FDR treatment of $\mu^2$~\cite{Pittau:2012zd,Donati:2013iya}: for consistency with eq.~(\ref{eq:deform}), a $q^2$ appearing in the numerator of a diagram should also be deformed 
\bqa
q^2 \to \qbar^2\,,
\eqa
and integrals involving $\mu^2$, such as
\bqa
{\tilde B}(p_1^2,M_0^2,M_1^2) = \int [d^4q] \frac{\mu^2}{\dbar_0 \dbar_1}\,,
\eqa
require the same integrand expansion {\em as if} $\mu^2= q^2$. 
For example, from  eq.~(\ref{eq:fdrquadtens}),
\bqa
{\tilde B}(p_1^2,M_0^2,M_1^2) &=& \lim_{\mu \to 0} \int d^4q \,
\mu^2 \left(
 \frac{d_1^3}{\qbar^8 \dbar_1}
+ \frac{d_0 d_1}{\qbar^6 \dbar_1}
+ \frac{d_0^2}{\qbar^4 \dbar_0\dbar_1}
\right)\nl
 &=& \frac{i \pi^2}{2}\left(M_0^2+M_1^2-\frac{p_1^2}{3}\right)\,.
\eqa
The final result follows by inserting eq.~(\ref{eq:cs}) into~(\ref{eq:gammav1})
\bqa
\label{eq:totvirt}
\Gamma_{V}(H \to gg)=\frac{3}{2} \frac{\als}{\pi}\,\Gamma^{(0)}(\als)\,
\left(\pi^2-\ln^2\frac{M^2_H}{\mu^2} 
\right)\,.
\eqa

\section{The real radiation $\Gamma_{R}(H \to ggg)$ and the fully inclusive result}
\label{realtot}
The unpolarized matrix element squared, derived from the real emission diagrams in Figure~\ref{fig:fig1}, reads
\bqa
|M|^2 &=& 192\,\pi \als A^2 \Bigg[\frac{s^3_{23}}{s_{12}s_{13}}
+\frac{s^3_{13}}{s_{12}s_{23}}
+\frac{s^3_{12}}{s_{13}s_{23}} 
+ \frac{2(s^2_{13}+s^2_{23})+3 s_{13}s_{23}}{s_{12}} \nl
&&+ \frac{2(s^2_{12}+s^2_{23})+3 s_{12}s_{23}}{s_{13}}
+ \frac{2(s^2_{12}+s^2_{13})+3 s_{12}s_{13}}{s_{23}} \nl
&& + 6 (s_{12} + s_{13} + s_{23})\Bigg]\,,
\eqa  
where $s_{ij}= (p_i+ p_j)^2$.
This expression is obtained from the massless result with
the replacement $2(p_i \cdot p_j) \to s_{ij}$, in accordance with eq.~(\ref{eq:repl}). 
As described in Section~\ref{fdr}, in order to match the virtual IR/CL singularities, $|M|^2$ should be integrated over a massive three-gluon phase space with $p_i^2= \mu^2$, which can be parametrized as
\begin{equation}
\int d \Phi_3 = \frac{\pi^2}{4s} \int ds_{12} ds_{13} ds_{23}\,
\delta(s-s_{12}-s_{13}-s_{23}+3\mu^2)\,,
\end{equation}
where $\sqrt{s}$ is the Higgs mass.
It is convenient to introduce the dimensionless variables
\begin{equation}
x= \frac{s_{12}}{s}-\mu_0\,,~~y= \frac{s_{13}}{s}-\mu_0\,,~~z= \frac{s_{23}}{s}-\mu_0\,,
\end{equation}
with $\mu_0$ given in eq.~(\ref{eq:mu0}), in terms of which, by using the condition
\bqa
x+y+z = 1\,, 
\eqa
all IR/CL divergent bremsstrahlung integrals can be reduced to
the following ones
\bqa
\label{eq:set}
I(s)    = \int_{R} dx dy\, \frac{1}{(x+\mu_0) (y+\mu_0)}\,,~\,
J_{p}(s) = \int_{R} dx dy\, \frac{x^p}{(y+\mu_0)}\,\,~~(p \geq 0)\,,
\eqa
where the integration region reads, in the fully inclusive case,
\bqa
\int_{R} dx dy \equiv 
\int_{3 \mu_0}^{1-2\sqrt{\mu_0}} dx \int_{y_-}^{y_+} dy\,,
\eqa
with
\bqa
y_{\pm} &=& \frac{1}{4(x + \mu_0)}
\left[(1-\mu_0)^2-(R_0 \mp R_1)^2\right] -\mu_0\,, \nl
R_0 &=& \sqrt{(x-\mu_0)^2-4 \mu_0^2}\,,~~~R_1 ~=~ \sqrt{(1-x)^2-4 \mu_0} \,.
\eqa
Thus
\bqa
\Gamma_{R}(H \to ggg)=
3 \frac{\als}{\pi}\,\Gamma^{(0)}(\als)\,
\left(\frac{1}{4}+I(M^2_H)-\frac{3}{2}J_0(M^2_H)-J_2(M^2_H)
\right)\,.
\eqa
Finally, one computes, up to terms which vanish in the limit $\mu_0 \to 0$,
\bqa
\label{eq:is}
I(s)      &=&  \frac{\ln^2(\mu_0)-\pi^2}{2}\,, 
\eqa
and
\bqa
\label{eq:icp}
J_{p}(s)    &=& -\frac{1}{p+1} \ln(\mu_0) 
+ \int_0^1 dx \,x^p\left[\ln(x)+ 2 \ln(1-x) \right] \nl
             &=& -\frac{1}{p+1} \ln(\mu_0)
-\frac{1}{p+1}\left[\frac{1}{p+1}+2 \sum_{n=1}^{p+1}\frac{1}{n} 
\right]\,\,~~(p \geq 0)\,,
\eqa
so that
\bqa
\Gamma_{R}(H \to ggg)=
\frac{3}{2} \frac{\als}{\pi}\,\Gamma^{(0)}(\als)\,
\left(\ln^2\frac{M^2_H}{\mu^2}-\pi^2+\frac{73}{6}
-\frac{11}{3}\ln \frac{M^2_H}{\mu^2}
\right)\,.
\eqa
Summing this to eq.~(\ref{eq:totvirt}), and adding the finite renormalization term in eq.~(\ref{eq:A}), one obtains
\bqa
\label{eq:eq1bis}
\Gamma(H \to gg) &=& \Gamma^{(0)}(\als)
\left[ 
1+\frac{\als}{\pi} 
\left(
\frac{95}{4}-\frac{11}{2} \ln\frac{M^2_H}{\mu^2}
\right)
\right]\,.
\eqa
All CL/IR $\ln(\mu^2)$ and $\ln^2(\mu^2)$ cancel in eq.~(\ref{eq:eq1bis}), so that the remaining $\mu$ is directly interpreted as the renormalization scale.
This is a typical procedure in FDR: since the UV infinities are subtracted from the very beginning, the unphysical left over $\mu$ dependence is eliminated, at the perturbative level one is working, by a finite renormalization, which fixes the bare parameters in terms of observables~\cite{Pittau:2013ica}. This is obtained, in the case at hand, by simply replacing $\Gamma^{(0)}(\als) \to \Gamma^{(0)}(\als(\mu^2))$~\footnote{$\als(\mu^2)$ has to be computed in the $\overline{\rm MS}$ scheme, as explained in the next Section.} in eq.~(\ref{eq:eq1bis}). Then, the logarithm is reabsorbed in the gluonic running of the strong coupling constant
\bqa
\als(M_H^2)= \frac{\als(\mu^2)}{1+\frac{\als}{2 \pi}\frac{11}{2} 
\ln\frac{M_H^2}{\mu^2}}\,,
\eqa
and eq.~(\ref{eq:eq1}) follows.

\section{FDR vs DR}
\label{fdrvsdr}
In this section, I discuss the transition rules between FDR and DR. This is particularly important in QCD, where NLO calculations have to be matched with the runnings of $\als$ and parton densities, conventionally derived in DR. I consider UV, CL and IR divergences in turn, showing the equivalence of FDR with the Dimensional Reduction~\cite{Siegel:1979wq} version of DR, widely used in supersymmetric theories. 

 I start by establishing the connection between the $1/\epsilon$ 
DR regulator and the $\ln(\mu^2)$ appearing in FDR.
As for the UV infinities, it is sufficient to compare the FDR and DR variants of any divergent integral. For instance, the DR counterpart of eq.~(\ref{eq:bp2}) (with $p^2 \neq 0$) reads
\bqa
\int d^nq \frac{1}{q^2(q+p)^2}= i \pi^2 \int_{0}^{1} dx\,
[\Delta -\ln(-p^2 x (1-x))]\,, 
\eqa
where
\bqa
n &=& 4 + \epsilon\,~~~{\rm and}~~~\Delta= -\frac{2}{\epsilon} - \gamma_E -\ln \pi\,.
\eqa
Thus, DR and FDR UV regulators are linked through the simple $\overline{\rm MS}$ replacement
\bqa
\label{eq:trans}
\Delta \rightarrow \ln(\mu^2)\,.
\eqa

CL virtual singularities follow the same pattern, as can be inferred from the exact UV/CL cancellation in eqs.~(\ref{eq:b1}) and~(\ref{eq:b2}).
As a consistency check, the DR version of $J_{p}$ reads
\bqa
J^{\rm DR}_{p}(s)  &=&  
\frac{(\pi s)^{\frac{\epsilon}{2}}}{\Gamma\left(1+\frac{\epsilon}{2}\right)}
\int dx\,dy\,dz\,\frac{x^p}{y}\delta(1-x-y-z) (xyz)^{\frac{\epsilon}{2}} \nl
             &=& -\frac{1}{p+1} (\Delta-\ln(s))
-\frac{1}{p+1}\left[\frac{1}{p+1}+2 \sum_{n=1}^{p+1}\frac{1}{n} 
\right]\,,
\eqa
which indeed coincides with eq.~(\ref{eq:icp}) if $\Delta= \ln(\mu^2)$.

Finally, the $\ln^2(\mu^2)$ terms -generated by overlapping IR/CL singularities- drop, together with the full constant part, when adding virtual and fully inclusive real contributions, which can be traced back to
the following relation 
\bqa 
{\cal R}e \left[\frac{C(s)}{i \pi^2}\right]= \frac{1}{s} I(s)\,
\eqa
between eqs.~(\ref{eq:cs}) and (\ref{eq:is}).
An easy calculation shows that the same happens in DR. In fact
\bqa
{\cal R}e \left[ \frac{1}{i \pi^2} \int d^nq \frac{1}{q^2(q+p_1)^2(q+p_2)^2} \right] = \frac{1}{s} (\pi s)^{\frac{\epsilon}{2}}\, \Gamma\left(1-\frac{\epsilon}{2}\right)
\left[
\frac{4}{\epsilon^2}-\frac{2}{3} \pi^2
\right]\,, 
\eqa
where $p_1^2= p_2^2= 0$ and $s= -2(p_1\cdot p_2)$, and
\bqa
I^{\rm DR}(s) &=& 
\frac{(\pi s)^{\frac{\epsilon}{2}}}{\Gamma\left(1+\frac{\epsilon}{2}\right)}
\int dx\,dy\,dz\,\frac{1}{xy}\delta(1-x-y-z) (xyz)^{\frac{\epsilon}{2}} \nl
&=& (\pi s)^{\frac{\epsilon}{2}}\,\Gamma\left(1-\frac{\epsilon}{2}\right)
\left[
\frac{4}{\epsilon^2}-\frac{2}{3} \pi^2
\right]\,. 
\eqa

In summary, eq.~(\ref{eq:trans}) is the only needed relation between the two regulators. However, an important difference between DR and FDR follows on from self contractions of metric tensors coming from the Feynman rules.
In DR $g_{\alpha\beta}g^{\alpha\beta}= n$, while $g_{\alpha\beta}g^{\alpha\beta}= 4$
in FDR. This, together with eq.~(\ref{eq:trans}), and the FDR treatment of $\mu^2$ discussed in section~\ref{virtual}, makes explicit the equivalence between FDR and Dimensional Reduction in the $\overline{\rm MS}$ scheme.
Having established this, all the well known transition 
rules between Dimensional Reduction and DR~\cite{signer,Pittau:2011qp} can be directly applied to FDR.
In the case of eq.~(\ref{eq:eq1bis}), it turns out that the expression is the same in both Dimensional Reduction (or FDR) and DR. Therefore the correct strong coupling constant to be used is the customary $\als(\mu^2)$ in the $\overline {\rm MS}$ scheme, proving that the FDR result coincides with eq.~(\ref{eq:eq1}).
\section{Conclusions}
\label{conc}
I presented a FDR calculation of the gluonic QCD corrections  to $H \to gg$ in the large top effective theory, demonstrating that ultraviolet, collinear and infrared divergences can be simultaneously and successfully regulated in four dimensions.
I proved the equivalence, at the one-loop level, of Dimensional Reduction and FDR, making the latter approach attractive also in supersymmetric calculations, where the fermionic and bosonic sectors must share the same number of degrees of freedom.

The advantage of directly working in the four dimensional Minkowsky space is expected to lead to considerable simplifications in higher order QFT computations, especially in connection with numerical techniques.
This issue, together with the extension of FDR to more loops, is currently under study.


\section*{Acknowledgments}
I thank Fabio Maltoni for pointing me out $H \to gg$ as a possible
case study of FDR applied to massless QCD.
This work was performed in the framework of the ERC grant 291377, ``LHCtheory: Theoretical predictions and analyses of LHC physics: advancing the precision frontier''. I also thank the support of the MICINN project FPA2011-22398 (LHC@NLO) and the Junta de Andalucia project P10-FQM-6552.



\bibliographystyle{paper}
\bibliography{paper}







\end{document}